\documentclass[%
 reprint,
superscriptaddress,
nofootinbib,
 showframe,
 amsmath,amssymb,
 aps,
 prd,
]{revtex4-2}

\usepackage{graphicx}
\usepackage{dcolumn}
\usepackage{bm}
\usepackage{hyperref}
\usepackage{xcolor}

\begin{document}

\preprint{APS/123-QED}

\title{Emulation of SPHEREx Galaxy Power Spectra I: Neural Network Details and Optimization}

\author{Joseph Adamo}
    \email{jadamo@arizona.edu}
    \affiliation{Department of Astronomy/Steward Observatory, University of Arizona, 933 North Cherry Avenue, Tucson, AZ 85721, USA\\}
\author{Grace Gibbins}
    \email{gibbins@arizona.edu}
    \affiliation{Department of Physics, University of Arizona, 1118 E Fourth Str, Tucson, AZ 85721, USA\\}
\author{Anne Moore}
    \email{amoore8@arizona.edu}
    \affiliation{Department of Astronomy/Steward Observatory, University of Arizona, 933 North Cherry Avenue, Tucson, AZ 85721, USA\\}
\author{Tim Eifler}
    \affiliation{Department of Astronomy/Steward Observatory, University of Arizona, 933 North Cherry Avenue, Tucson, AZ 85721, USA\\}
    \affiliation{Department of Physics, University of Arizona, 1118 E Fourth Str, Tucson, AZ 85721, USA\\}

\date{\today}

\begin{abstract}
We present neural networks to generate redshift-space galaxy power spectrum multipoles for multiple tracer and redshift bins simultaneously given a set of input cosmology and galaxy bias parameters. This emulator utilizes a combination of fully-connected layers and transformer architecture to accurately predict galaxy power spectrum multipoles $900$ times faster than the SPHEREx pipeline. We quantify network performance using both $\Delta \chi^2$, and likelihood contours for simulated SPHEREx analyses, using two correlated tracer bins and two independent redshift bins. After optimizing network architecture, the loss function, and training set sampling strategy, we achieve $\operatorname{Med}\left( \Delta \chi^2\right) = 0.069$ when comparing to our testing set. At the contour-level our emulator  agrees with EFT predictions over a realistic parameter range, with an average 1D best-fit shift of $0.078\sigma$ and $0.82 \%$ change in 1D error bars. These results demonstrate the feasibility of using neural-network emulators to accelerate SPHEREx redshift-space power-spectrum analyses.

\end{abstract}

\maketitle


\section{Introduction}
\label{sec:intro}



Cosmologists have access to an ever-increasing library of data. Currently, stage III missions like the Dark Energy Survey (DES \cite{DES-Y6-Catalog}), the Hyper Suprime Camera (HSC, \cite{HSC-Y3-data}), and the Atacama Cosmology Telescope (ACT, \cite{ACT-DR6-Data}) are at or near completion, while several stage IV missions like the Dark Energy Spectroscopic Instrument (DESI, \cite{DESI-Y2-BAO}), and Euclid \cite{Euclid-Early-Data} have already begun releasing early results. This trend will only continue with new surveys coming online, such as Rubin Observatory's Legacy Survey of Space and Time\footnote{\url{https://rubinobservatory.org/}} (LSST, \cite{Rubin-LSST}), and the Nancy Grace Roman Space Telescope\footnote{\url{https://roman.gsfc.nasa.gov/}}.

One such new mission is the Spectro-Photometer for the History of the Universe, Epoch of Reionization and Ices Explorer (SPHEREx), which launched in early March 2025. SPHEREx is probing the large-scale structure (LSS) of the late universe by conducting an all-sky near-infrared $(0.75 < \lambda < 5\ \mu m)$ spectroscopic galaxy survey \citep{SPHEREx-preliminary-strategy}. Over the course of two years, we expect to receive precise redshift measurements for $\sim 450$ million galaxies, which will allow us to precisely estimate the galaxy power spectrum and bispectrum at very large scales \cite{SPHEREx-status}.

Analyzing these summary statistics will place tight constraints on early-universe inflation by measuring imprints of local-type primordial non-Gaussianity (PNG) on the late-time matter distribution through the parameter $f^{\rm{loc}}_{\rm{NL}}$ (henceforth referred to as simply $f_{\rm{NL}}$). This parameter is typically $\sim0.01$ in single-field inflation models and $\gtrsim  1$ in multi-field theories \citep{Primordial-non-Gaussianity, Inlfation-fnl-predictions}. Currently, our tightest constraints come from Planck, with $f_{\rm{NL}} = -0.9\pm 5.1$ \cite{Plack-fnl}. Recent full-shape analyses have found $f_{\rm{NL}} = -33 \pm 28$ using BOSS galaxies \cite{BOSS-fnl-1}, and $-4 < f_{\rm{NL}} < 27$ with eBOSS quasars \cite{eBOSS-fnl-1, eBOSS-fnl-2}. Using the first year of DESI data, Refs. \cite{DESI-fnl-1, DESI-fnl-2} report $f_{\rm{NL}} = -3.6^{+9.0}_{-9.1}$ and $f_{\rm{NL}} = -0.1 \pm 7.4$, respectively. These galaxy clustering measurements are beginning to approach the precision of CMB constraints, and can be further improved by increasing the effective survey volume. Indeed, SPHEREx aims to measure PNG with a sensitivity of $\sigma (f_{\rm{NL}}) \sim 1$ as its primary cosmological science objective, allowing us to differentiate between the two inflation scenarios \citep{SPHEREx-cosmology}.

To prepare for constraining $f_{\rm{NL}}$ with SPHEREx, specifically with the power spectrum, we need to quantify how analysis choices impact the resulting constraints via simulated likelihood analyses. These choices include investigating the specific priors used for cosmology and galaxy bias parameters, what scales to include in the  analysis, and the impact of the covariance matrix, to name a few. However, performing these analyses involves predicting the galaxy power spectrum at $\sim 10^6$ points in parameter space to adequately sample the posterior distribution. Most modeling codes take $\sim 1\mathrm{s}$ per evaluation, making posterior sampling computationally prohibitive when varying a realistic number of parameters.

To enable fast evaluations during likelihood analyses, we present the Multipole Emulator for Nonlinear Tracer Analysis of Two-point statistics and Large Scale Structure (MENTAT-LSS), a fast method for emulating galaxy power spectrum multipoles using machine learning. We develop a set of deep neural networks to directly emulate the galaxy power spectrum for multiple tracer and redshift bins at once by using a combination of fully connected layers and transformer architecture \cite{Attention-paper}. Our emulator is trained on SPHEREx-like galaxy power spectrum multipoles predicted by Eulerian effective-field theory (EFT, \cite{CLASS-PT}). We evaluate performance w.r.t. network architecture, training choices, and the parameter-sampling used to construct the training set, and select a configuration that maximizes emulator accuracy. Finally, we use our emulator to run SPHEREx-like analyses within $\Lambda$CDM using two correlated tracer bins, and two independent redshift bins.

This paper will proceed as follows. Section \ref{sec:background} describes the galaxy power spectrum model we emulate, as well as how we generate a training set with said model. Section \ref{sec:architecture} discusses the neural network architecture and training scheme, while Section \ref{sec:testing} discusses both how the emulator is optimized and the performance metrics we use to quantify accuracy. Finally, we use our emulator in a simulated SPHEREx likelihood analysis in Section \ref{sec:analysis}, and summarize our findings in Section \ref{sec:conclusion}.

\section{Emulating the Galaxy Power Spectra}
\label{sec:background}

\subsection{EFT Model Details}

The Eulerian effective field theory (EFT) prediction for the redshift-space galaxy power spectrum has been used extensively to analyze BOSS data with both Gaussian initial conditions \cite{CLASS-PT, Ivanov-2020, Wadekar-2020, CovNet}, and local PNG \cite{BOSS-fnl-1}. More recently, it was also used to independently constrain both dark energy and PNG with DESI data \cite{Chudaykin-2025, DESI-fnl-2}. To summarize, this EFT model splits the full galaxy power spectrum into four components (omitting IR resummation and the Alcock-Paczynski effect in our expressions for simplicity),

\begin{eqnarray}
   \nonumber
    P^\text{EFT} (k, \mu) = \left(b_1 + f\mu^2\right)^2P_\text{lin}(k) + P_\text{1-loop} (k, \mu) + \\
    P_\text{ctr}(k, \mu) + P_\text{shot}(k, \mu) ,
\label{eq:eft}
\end{eqnarray}

\noindent where $f$ is the growth factor at the effective redshift of the given galaxy sample, $\mu$ is the cosine line-of-sight angle, and $b_1$ is linear galaxy bias. The first term in Eq. \ref{eq:eft} describes the linear Kaiser effect \cite{Kaiser-effect}, while the one-loop term 
corrects for nonlinear effects from both structure growth and redshift-space distortions (RSD). To calculate this term, we expand the galaxy density contrast\footnote{In this paper, we assume Gaussian initial conditions. When accounting for local PNG, this expansion includes several more bias terms that depend on $f_{\rm{NL}}$.} $\delta_g(\textbf{x})$ using the following basis,

\begin{equation}
    \delta_g(\textbf{x}) = b_1 \delta(\textbf{x}) + \frac{b_2}{2} \delta(\textbf{x})^2 + b_{\mathcal{G}_2} \mathcal{G}_2(\textbf{x}) \,
\label{eq:basis-1}
\end{equation}

\noindent where $\delta(\textbf{x})$ is the nonlinear matter density contrast, $\mathcal{G}_2$ is the tidal field operator in Fourier space, and $b_2$ and $b_{\mathcal{G}_2}$ are second-order bias parameters. Following the baseline DESI collaboration full-shape analysis choices, we set all third-order bias terms to $0$ for this work \cite{DESI-Y1-Fullshape}. Next, $P_\text{ctr}(k, \mu)$ corrects for small-scale physics integrated over by the one-loop term \cite{Senatore-2014}. Following the literature, we include two multipole-specific terms and one next-to-leading order term with the following form,

\begin{eqnarray}
    \nonumber
    P_\text{ctr}(k, \mu) = -2 c_0 k^2 P_\text{lin}(k) - 2 c_2 k^2  P_\text{lin}(k) - \\
    \tilde{c} f^4 k^4 (b_1 + f\mu^2)P_\text{lin}(k) ,
\end{eqnarray}

\noindent where $c_0, c_2, \tilde{c}$ are free parameters. This term is cheap to calculate provided one has access to the linear power spectrum. Thus, we do not directly emulate $P_{ctr}(k, \mu)$, and instead calculate it normally after emulating the tree and one-loop terms.

Finally, $P_\text{shot}$ accounts for stochastic contributions to the power spectrum. Similar to the counterterms, we do not emulate this term directly. In total, our EFT model contains three emulated nuisance parameters ($b_1, b_2, b_{\mathcal{G}_2}$), and four analytically calculated parameters ($c_0, c_2, \tilde{c}, P_{shot}$). When adding IR resummation and the Alcock-Paczynski effect, this model produces accurate predictions up to $k = 0.3$ h/Mpc \cite{CLASS-PT}.

\subsection{Training Data}

We generate training sets based on the EFT model described above. Specifically, we use a pre-release version of the EFT power spectrum code \texttt{ps\_1loop} \cite{ps_1loop} to compute galaxy power-spectrum  $\ell=(0,2)$ multipoles including tree-level and one-loop contributions. This code is similar to \textit{CLASS-PT} \cite{CLASS-PT}, but also accounts for local PNG using the formalism in Ref. \cite{BOSS-fnl-1}. We generate four million model evaluations using embarrassingly parallel execution. For each set of input cosmology and bias parameters, we output multipoles for two independent redshift bins ($0.2 < z < 0.4, 0.4 < z < 0.6$), and two correlated tracer bins with $\frac{\sigma(z)}{1+z} < (0.003, 0.1)$. These specific sample definitions are taken from current SPHEREx public products that also provide galaxy number densities, and linear bias values for each bin\footnote{\url{https://github.com/SPHEREx/Public-products/blob/master/galaxy_density_v28_base_cbe.txt}}. For the power spectrum, we opt to use twenty-five k-bins spanning $k_{min} = 0.001$ to $k_{max} = 0.2$ h/Mpc with uniform bin width in k.

We generate a training set with the above configurations by sampling parameters uniformly within a hypersphere \cite{hypersphere-sampling}. Hypersphere sampling effectively cuts out low-probability regions of parameter space, which leads to an increased density of points within the range of interest compared to a Latin hypercube. We compare this sampling strategy to a training set generated with a Latin hypercube in Section \ref{subsec:training-sampling}. Both training sets use the parameter ranges listed in Table \ref{tab:priors}, where $80\%$ of the full dataset is used for updating network weights during training, $10\%$ for validation during training, and $10\%$ for evaluating the final model performance.

To facilitate the emulator training process, we normalize all input parameters to lie within the range $[0, 1]$. Additionally, we further post-process our training set in a similar fashion as Refs. \cite{LSST-emulator-1, LSST-emulator-2}. Given a data covariance matrix, we first compute its eigenvalue decomposition,

\begin{equation}
    C = Q \Sigma Q^{-1}.
\end{equation}

\noindent We obtain $C$ via a modified version of CovaPT \cite{CovaPT} configured to output multi-tracer Gaussian covariance matrices. Next, we transform each set of power spectrum multipoles to lie within the diagonal basis defined by the eigenvectors $Q$ using the following equation,

\begin{equation}
    \hat{P}_i = \frac{Q^{-1}P_i - Q^{-1} P_{fid}}{\sqrt{\Sigma}}.
\end{equation}

\noindent Here, $P_{fid}$ are the galaxy power spectrum monopole and quadropole calculated using the fiducial cosmology in Table \ref{tab:priors}. Our emulator is thus trained to output a whitened version of the data $\hat{P}$, which is then post-processed back into galaxy power spectrum multipoles. We note that this method differs from compressing the data using principle component analysis (PCA) used throughout the field (e.g. Refs. \cite{Derose-2022, COSMOPOWER, PCA-example-1}). Rather than performing dimensionality reduction, our transformation preserves the full information content while improving training stability.

\begin{table}[b]
    \centering
    \begin{tabular}{ccc}
    \hline
    Parameter & Fiducial Value & Emulator Range\\
    \hline \hline
    h & $0.6736$ & $U[0.44, 0.9]$\\
    $\omega_{cdm}$ & $0.1201$ & $U[0.0701, 0.1701]$ \\
    $\omega_b$ & $0.02218$ & $U(0.01998, 0.02438]$ \\
    $A_s$ & $2.1 \times 10^{-9}$ & $U[1.2 \times 10^{-9}, 3 \times 10^{-9}]$ \\
    $n_s$ & $0.96589$ & $U[0.83989, 1.09189]$\\
    \hline
    $b_1^{(0,0)}$ & $1.5$ & $U[0.25, 2.75]$ \\
    $b_1^{(0,1)}$ & $1.3$ & $U[0.05, 2.55]$ \\
    $b_1^{(1,0)}$ & $1.8$ & $U[0.55, 3.05]$ \\
    $b_1^{(1,1)}$ & $1.4$ & $U[0.15, 2.65]$ \\
    $b_2^{(0,0)}$ & $-1.37$ & $U[-3.12, 0.38]$ \\
    $b_2^{(0,1)}$ & $-1.57$ & $U[-3.32, 0.18]$ \\
    $b_2^{(1,0)}$ & $-0.78$ & $U[-2.53, 0.97]$ \\
    $b_2^{(1,1)}$ & $-1.49$ & $U[-3.24, 0.26]$ \\
    $b_{\mathcal{G}_2}^{(0,0)}$ & $-0.29$ & $U[-1.79, 1.21]$ \\
    $b_{\mathcal{G}_2}^{(0,1)}$ & $-0.17$ & $U[-1.67, 1.33]$ \\
    $b_{\mathcal{G}_2}^{(1,0)}$ & $-0.46$ & $U[-1.96, 1.04]$ \\
    $b_{\mathcal{G}_2}^{(1,1)}$ & $-0.23$ & $U[-1.73, 1.27]$ \\
    \hline\hline
    
    \end{tabular}
    \caption{Table describing the fiducial cosmology and parameter bounds used in our training sets and simulated analyses. For cosmology parameters, we adapt the priors and fiducial values from the baseline analysis in Ref. \cite{DESI-Y1-Fullshape}. For galaxy bias, we set different fiducial values for each tracer and redshift bin for $b_1$, and estimate the fiducial values for $b_2$ and $b_{\mathcal{G}_2}$ using the fits from Refs. \cite{Baldauf-2012, Lazeyras-2016}. We set all other bias parameters to 0 unless otherwise specified.}
    \label{tab:priors}
\end{table}

\begin{figure*}
    \centering
    \includegraphics[width=0.8\linewidth]{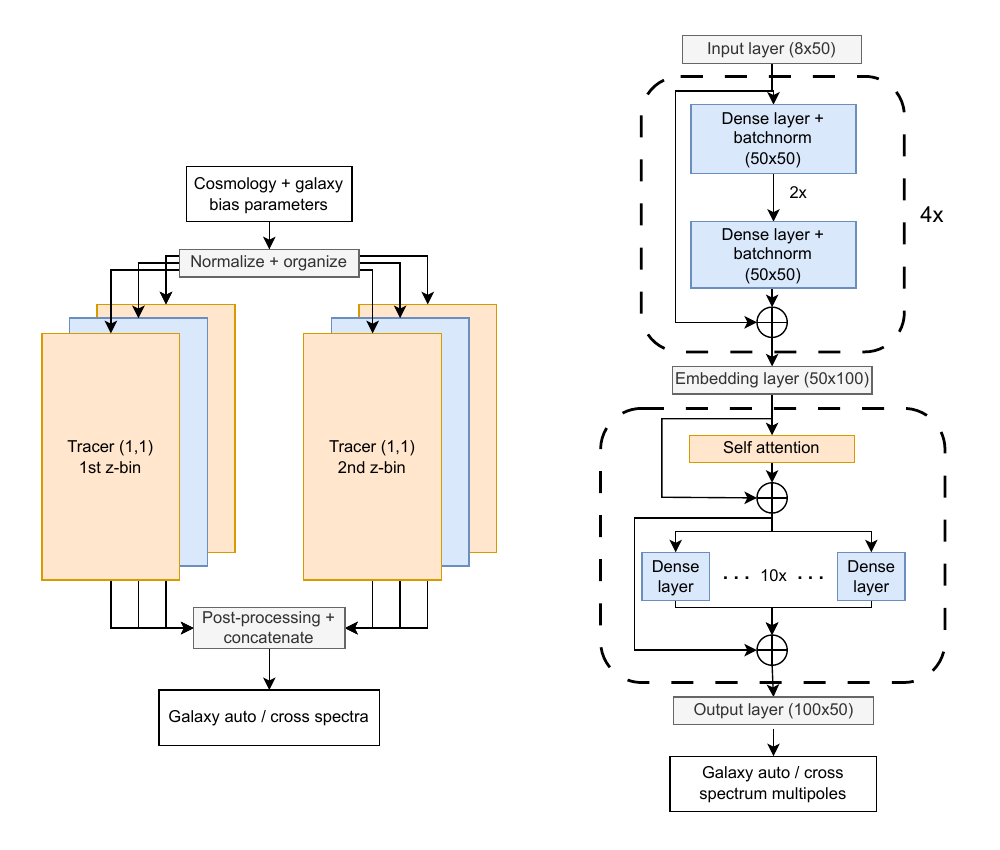}
    \caption{(Left): Schematic of our full galaxy power spectrum emulator. Cosmology and galaxy bias parameters are normalized and passed to six independent neural networks, each responsible for generating the auto or cross power spectrum for a particular tracer and redshift bin. We then concatenate the output from these networks to output the galaxy power spectrum for two tracer and two redshift bins.\\
    (Right): Architecture of an individual network in the full emulator. Relevant parameters first pass through a series of four MLP blocks, followed by an embedding layer. After that, we proceed through one transformer encoder block using the same architecture as Ref. \cite{LSST-emulator-2}, before finally outputting normalized multipoles for a specific redshift and tracer bin.}
    \label{fig:network_diagram}
\end{figure*}

\section{Neural Network Architecture}
\label{sec:architecture}

Our emulator assigns independent networks to each set of (auto or cross) power spectrum multipoles corresponding to an individual tracer and redshift bin. Since we are correlating different tracers within a given redshift bin, we have three networks per redshift bin (two auto, one cross spectra) for our chosen configuration. This design choice allows us to keep each network relatively small, making the emulator easier to train and expand to larger tracer / redshift bin setups. Each network is sensitive to all cosmology but only the relevant galaxy bias parameters for that particular bin. We provide a schematic of our emulator in Fig. \ref{fig:network_diagram}.

The networks themselves follow a similar design to the complete attention-based architecture from Ref. \cite{LSST-emulator-2}, which we summarize here. The (normalized) cosmology and bias parameters first pass through an initial input layer, then through a series of four fully-connected multi-layered perceptron (MLP) blocks. Each of these blocks has two MLP layers with batch normalization, followed by a residual connection. The output of those blocks gets sent to an embedding layer that doubles the dimensionality to one hundred, before passing through one transformer block. In this block, the input goes through a scaled dot-product attention mechanism, and is then split into ten equally-sized components that are each fed to independent feed-forward layers. Finally, the transformer block output passes through a final MLP layer, which outputs the galaxy power spectrum monopole and quadropole moments for a specific tracer and redshift bin combination. We chose the specific number of MLP and transformer blocks, as well as the number of layers per MLP block, based on the optimization results in Section \ref{subsec:optimization} .

After each layer within the MLP blocks, and the feed-forward layer in the transformer blocks, we use the following nonlinear activation function,

\begin{equation}
    h(x) = \left(\gamma + \left(1 + e^{-\beta \odot x}\right)^{-1} \odot (1 - \gamma)\right) \odot x,
\end{equation}

\noindent where $\beta, \gamma$ are trainable parameters of the model with the same shape as the input vector $x$. This function is thought to more accurately capture both sharp and smooth features due to its flexibility, and has been used by other emulators in the field \cite{COSMOPOWER, CONNECT, LSST-emulator-2}.

Each network is trained separately by stochastic gradient descent. We tested several different loss functions to minimize, all of which are based on the $\Delta \chi^2$ between the emulator output and the prediction from our EFT model,

\begin{align}
    \Delta \chi_{(z, tt)}^2 = \left(P_{(z, tt)}^{\rm emu} - P_{(z, tt)}^{\rm EFT}\right)^T C_{(z, tt)}^{-1} \left(P_{(z, tt)}^{\rm emu} - P_{(z, tt)}^{\rm EFT}\right).
    \label{eq:delta_chi2}
\end{align}

\noindent Here, $z$ denotes the specific redshift bin, $tt$ describes the specific tracer combination (in our case, 11, 12, or 22), and $C$ is the data covariance matrix for that combination. When evaluating accuracy after training, we use the $\Delta \chi^2$ calculated from all tracer and redshift bins as follows,

\begin{align}
    \Delta \chi^2_{\rm tot} = \left(\mathbf{P}^{\rm emu}-\mathbf{P}^{\rm EFT}\right)^{\!T} \mathbf{C}^{-1} \left(\mathbf{P}^{\rm emu}-\mathbf{P}^{\rm EFT}\right),
\label{eq:delta_chi2_tot}
\end{align}

\noindent where $\mathbf{P}$ denotes the data vector formed by concatenating $\mathbf{P}_{(z,tt)}$ across all tracer combinations and redshift bins, and $\mathbf{C}$ is the full multi-tracer covariance matrix. We describe the specific loss functions we tested in Section \ref{subsec:optimization}.

We utilize the Adam optimization scheme for updating our network weights \cite{Adam-optimization}. To allow for adequate loss function exploration without preventing the networks from reaching a minimum, we also implement an adaptive learning rate scheduler that reduces the learning rate by a factor of ten if the validation set loss does not improve after fifteen epochs. Finally, to prevent overfitting we use early stopping, where training ends after the validation loss has not improved after twenty-five epochs. 

\section{Optimization and Testing}
\label{sec:testing}

\subsection{Hyperparameter Optimization}
\label{subsec:optimization}

We optimize our emulator by testing different hyper-parameter settings, To quantify their impact on model performance, we train a series of emulators while individually varying the initial learning rate, batch size, number of MLP blocks, number of layers per MLP block, and number of transformer blocks. The list of values  iterated through is given in Table \ref{tab:hyperp}. The default parameter values were chosen based on Ref. \cite{LSST-emulator-2}, which uses a similar network design to emulate LSST observables. We use the median $\Delta \chi_{tot}^2$ calculated from the test set using equation \ref{eq:delta_chi2} as our main performance metric.

\begin{table*}[t]
\centering
\begin{tabular}{cccc}
 \hline
 Hyperparameter & Default Value & Iteration Values & Final Value \\ [0.5ex] 
 \hline\hline
 Learning Rate & 0.001 & 0.0001, 0.0005, 0.0025, 0.005, 0.0075, 0.01 & 0.005\\ 
 \hline
 Batch Size & 200 & 50, 100, 300, 500 & 200\footnote{Set to non-optimal value after accounting for training time.} \\
 \hline
 MLP Blocks & 2 & 1, 3, 4, 5 & 4\\
 \hline
 Layers / Block & 4 & 2, 3, 5, 6 & 2\\
 \hline
 Transformer \\ Blocks & 1 & 2, 3, 4, 5 & 1\footnotemark[1] \\ [1ex] 
 \hline\hline 
\end{tabular}
\caption{Table of network hyperparameters varied in our optimization scheme. We train emulators for each iteration value, holding all other hyperparameters at their default values. We use $\operatorname{Med}\left( \Delta \chi_{tot}^2\right)$ to determine which configuration performs best, the results of which (after accounting for training time) are displayed in column $4$.}
\label{tab:hyperp}
\end{table*}

We also explore two different loss functions given by the following equations,

\begin{align}
    \mathcal{L}_1 = \langle \Delta \chi_{(i, tt)}^2 \rangle,
    \label{chi2}\\
    \mathcal{L}_2 = \langle \sqrt{1 + 2\Delta \chi_{(i, tt)}^2} \rangle - 1 \label{hchi2},
\end{align}

\noindent which we call $\chi^2$ and hyperbolic $\chi^2$ loss respectively. We expect $\mathcal{L}_1$ to be more sensitive to outliers than $\mathcal{L}_2$, as it depends on the difference between emulated and true power spectra squared. For each loss function, we iteratively vary the hyper-parameters in Table \ref{tab:hyperp} to account for potentially different optimal configurations.

Table \ref{tab:hyperp} shows our final hyperparameter setup. Notably, batch-size appears to have an inverse relationship with both performance and training time, with values lower than $200$ taking an unreasonably long time to finish training. Thus, we chose to adopt a final batch size of $200$ to keep our results more easily scalable to larger setups. Similarly, increasing the number of transformers marginally improves performance, but also significantly increases training time. Thus, for the same reasons as above, we opt to use one block as our final value.

Additionally, we find that the above conclusions hold regardless of the loss function used. In fact, neither loss function performs consistently better or worse than the other, with hyperbolic $\chi^2$ performing $0.5\%$ better when averaging over all our training runs. These differences may be accounted for by the training process being inherently stochastic, thus we conclude that both $\mathcal{L}_1$ or $\mathcal{L}_2$ perform the same for this problem. In what follows, we chose to use $\mathcal{L}_2$ as our ``optimal'' loss function.

\subsection{Training Set Optimization}
\label{subsec:training-sampling}

Next, we explore the impact of the training set by varying both the number of power spectra we use to train, and the method of sampling the input cosmology and bias parameters. Both tests are done using the optimal set of hyper-parameters found from our above efforts.

\begin{figure}
    \centering
    \includegraphics[width=\linewidth]{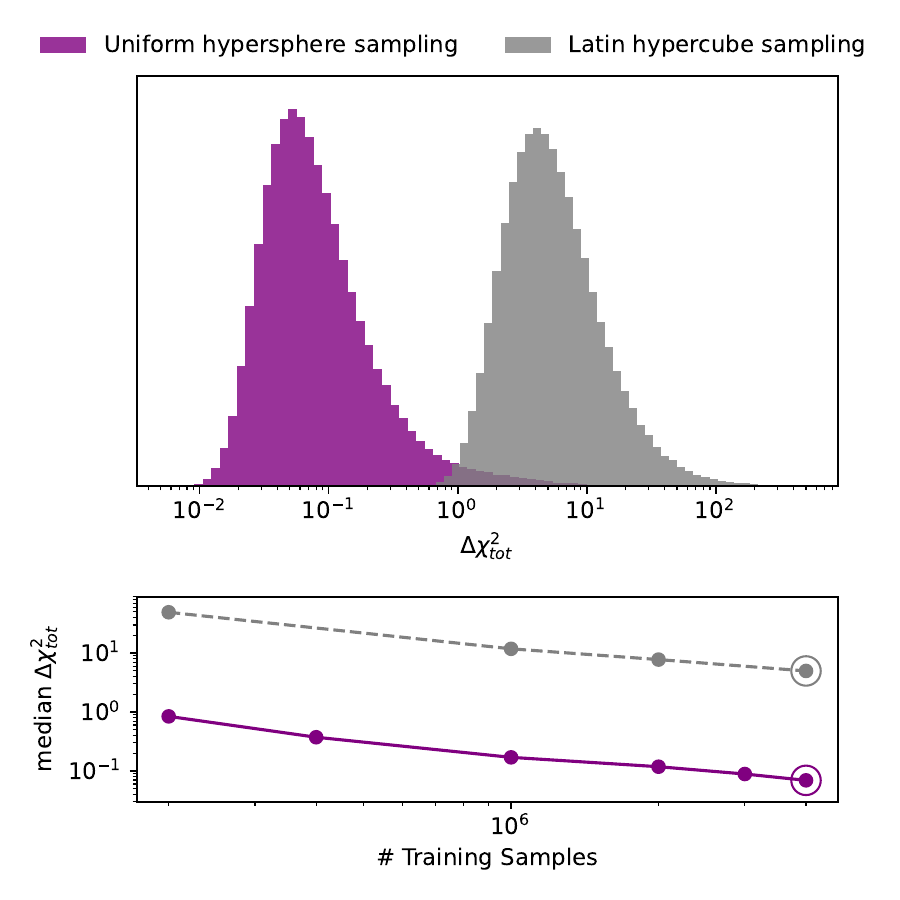}
    \caption{(Top): Histogram of $\Delta \chi_{tot}^2$ values for emulators trained on training sets sampled with a Latin hypercube (gray) and a uniform hypersphere (purple). We see two orders of magnitude difference between the two sampling strategies.\\
    (Bottom): Median $\Delta \chi_{tot}^2$ with respect to the total number of training samples for each sampling strategy. Circled points correspond to the histograms in the top panel. In both cases, performance improves roughly following a power law.}
    \label{fig:tset_size}
\end{figure}

The top panel of Fig. \ref{fig:tset_size} shows $\Delta \chi_{tot}^2$ histograms for emulators trained on differently-sampled training sets. We see that using hypersphere sampling results in two orders of magnitude improvement in performance compared to using a traditional Latin hypercube. Such a stark improvement can likely be attributed to the overall smaller volume and thus increased density given the same number of samples \cite{hypersphere-sampling}. While this improvement reduces global generalization, our analyses require high accuracy mainly over regions with non-negligible posterior weight. We can therefore center the hypersphere on a well-motivated reference cosmology (e.g., external best-fit constraints) and choose radii that conservatively enclose plausible posteriors. If samples approach the domain boundary, we can expand and/or re-center the training set and retrain.

Next, the bottom panel of Fig. \ref{fig:tset_size} shows how the median $\Delta \chi_{tot}^2$ changes with the number of training set samples. We find that both sampling strategies roughly follow a power law with the form,

\begin{equation}
    \operatorname{Med}\left(\Delta \chi_{tot}^2\right) = A(N_{sample})^{-x},
\end{equation}

\noindent with $x=(0.798, 0.771)$ for hypersphere and hypercube sampling respectively. Extrapolating outwards, we can naively expect performance to double after increasing the training set size by a factor of around $2.4$ in both cases. This result follows other scaling relations seen in a broad range of machine learning problems \cite{NN-scaling-1, NN-scaling-2}. Here, this relation means that our model could still achieve greater accuracy with larger training sets, which may become important when increasing the number of tracer and redshift bins.

\begin{figure}
    \centering
    \includegraphics[width=\linewidth]{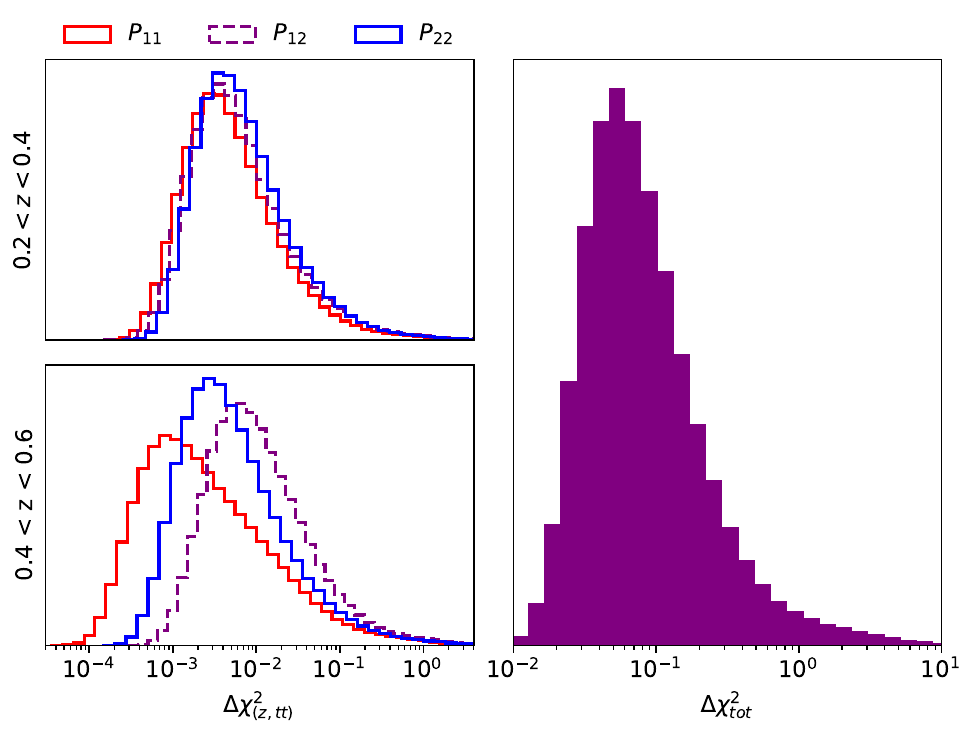}
    \caption{(Left): Histograms of $\Delta \chi_{(z, tt)}^2$ for each individual network in the final optimized emulator. Each panel denotes a specific redshift bin, and we denote auto/cross power spectra with solid and dashed lines respectively.\\
    (Right): Histogram of $\Delta \chi_{tot}^2$ calculated from the full optimized emulator output, which corresponds to the purple hypersphere-sampled histogram in Fig.\ref{fig:tset_size}.}
    \label{fig:chi2-histogram}
\end{figure}

Finally, we present more detailed $\Delta \chi^2$ statistics for our final optimized emulator. Figure \ref{fig:chi2-histogram} displays $\Delta \chi_{(z, tt)}^2$ histograms for each sub-network, and also for the full emulator output. We see that each network performs well in this metric, with most histograms peaking around $10^{-2}$. The higher redshift bin shows notably larger variation between networks, with $P_{(1, 11)}$ peaking at $10^{-3}$. Looking at the full output, we see a slightly higher distribution with $\operatorname{Med}\left( \Delta \chi_{tot}^2\right) = 0.069$, which passes the $\Delta \chi^2 < 1$ criterion used by both the DES collaboration, and other cosmological emulators  \cite{DES-Y3-Validation, LSST-emulator-1}. This increase is actually larger than summing all of the sub-network errors, due to including off-diagonal blocks of the covariance matrix that are not accounted for when calculating $\Delta \chi_{(z, tt)}^2$. As a consequence, we expect that increasing the number of tracer and redshift bins will result in larger $\Delta \chi_{tot}^2$, even if each sub-network's accuracy is kept the same as here.

\begin{figure*}
    \centering
    \includegraphics[width=\linewidth]{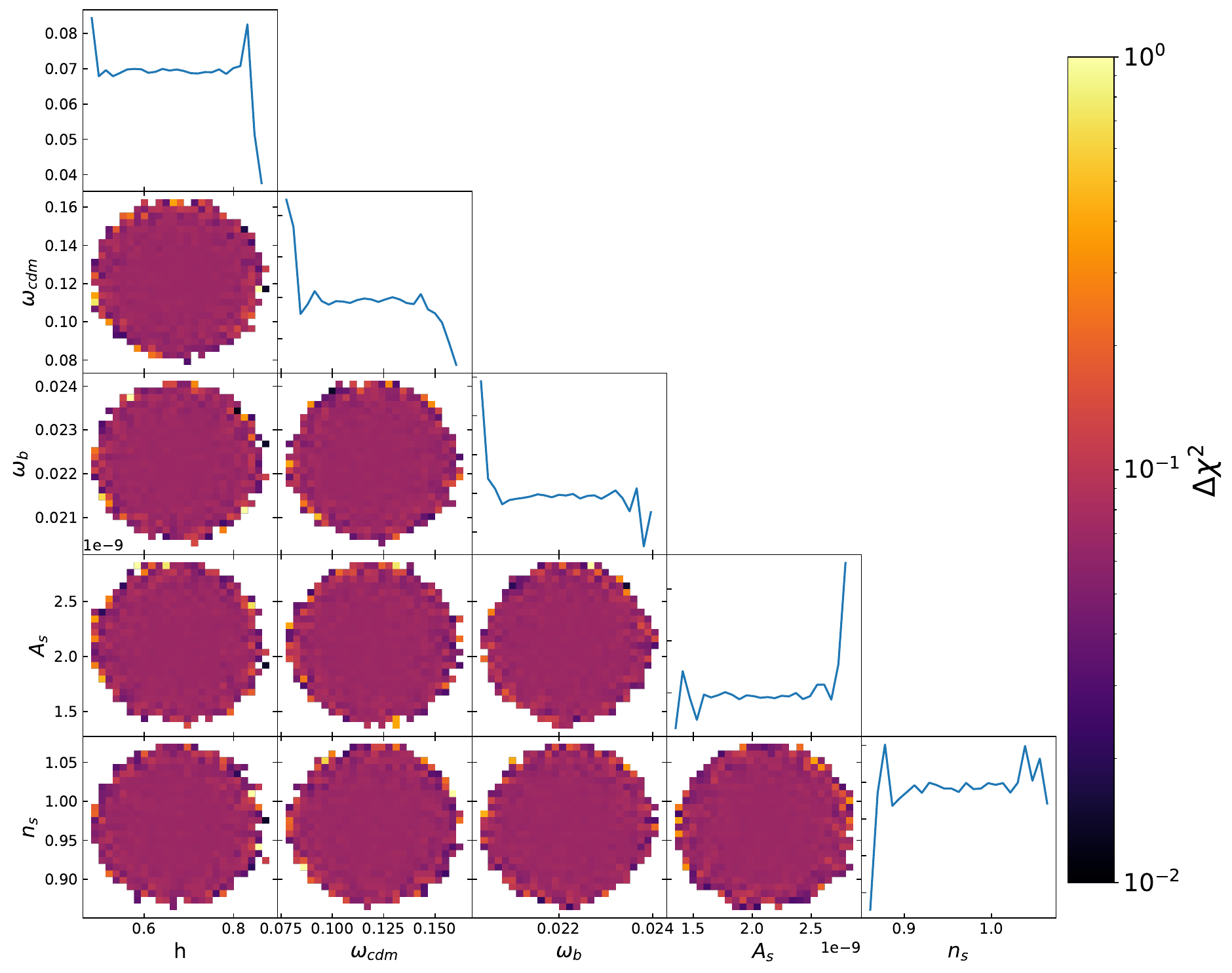}
    \caption{Heatmap showing the median $\Delta \chi_{tot}^2$ of our emulator throughout parameter space, Each pixel represents the median value from samples in the test set whose parameters fall in that specific range, represented via Eq. \ref{eq:heatmap_calc}. Other than some regions along the edges, the emulator performs similarly across the entirety of parameter space.}
    \label{fig:chi2-heatmap}
\end{figure*}

Figure \ref{fig:chi2-heatmap} shows the median $\Delta \chi_{tot}^2$ over the emulated parameter space. Specifically, each pixel is calculated by taking the median of all $\Delta \chi_{tot}^2$ values whose input parameters $\theta_x, \theta_y$ lie within the given (x,y) pixel via the following, 

\begin{equation}
    \Delta \chi^2_{\rm tot}(\theta_x,\theta_y) =
\operatorname{Med}\left[ \int
d\boldsymbol{\theta}'\,
\Delta\chi^2_{\rm tot}(\boldsymbol{\theta}) \right],
    \label{eq:heatmap_calc}
\end{equation}

\noindent where $\boldsymbol{\theta}' \equiv \boldsymbol{\theta} \setminus \{\theta_x, \theta_y \}$ refers to all parameters except $\theta_x, \theta_y$. This process effectively marginalizes over all other (cosmology and galaxy bias) parameters. We can see that performance is relatively stable over the full domain, with larger variation in some pixels along the edges. This trend indicates that we should not see any bias due to variations in performance when running simulated analyses, so long as said analyses does not hit the parameter domain boundary.

\section{Comparing Emulated vs. Direct EFT Likelihood Analyses}
\label{sec:analysis}

\subsection{Metric 1: Parameter Contours}

We run simulated analyses to showcase our optimized emulator by varying all cosmology and bias parameters listed in Table \ref{tab:priors}. Fiducial values for cosmology parameters are taken from the baseline DESI Y1 full modeling analyses \cite{DESI-Y1-Fullshape}. For galaxy bias, we assume different fiducial values of $b_1$ for each tracer and redshift bin, and calculate fiducial values for $b_2, b_{\mathcal{G}_2}$ using fits derived in Refs. \cite{Baldauf-2012, Lazeyras-2016}. We adopt uniform priors over the parameter ranges spanned by the emulator training set. The exceptions are $\omega_b$ and $n_s$, for which we impose Gaussian priors taken from the baseline DESI-Y1 full-modeling analysis \cite{DESI-Y1-Fullshape},

\begin{equation}
    \omega_b \sim \mathcal{N}(0.02218, 0.00055^2), n_s \sim \mathcal{N}(0.96589, 0.042^2).
\end{equation}

\noindent Since our emulator uses explicit ranges for these parameters, we also apply a cutoff with the ranges given in Table \ref{tab:priors}, corresponding to $\pm 4 \sigma, \pm 3 \sigma$ respectively. Finally, we neglect the effect of massive neutrinos for these analyses, so $m_\nu = 0$ eV.

For all analyses, we assume a Gaussian likelihood, with the same Gaussian covariance matrix used to train our emulator. We present the resulting marginalized contours in $[H_0,\Omega_m,\sigma_8]$ as a compact summary of the full posterior, since these parameters are both widely reported and provide an interpretable snapshot of the main cosmological constraints. Our chains are run using the \textit{Nautilus} python package \cite{Nautilus}, which estimates both the posterior distribution and evidence factor $\log \mathcal{Z}$ using importance-nested sampling. We use a stopping criteria of $n_\text{eff} = 10^5$ points to determine the chains are converged. Finally, we generate our contour plots using \textit{GetDist} \cite{GetDist}.

\begin{figure}
    \centering
    \includegraphics[width=\linewidth]{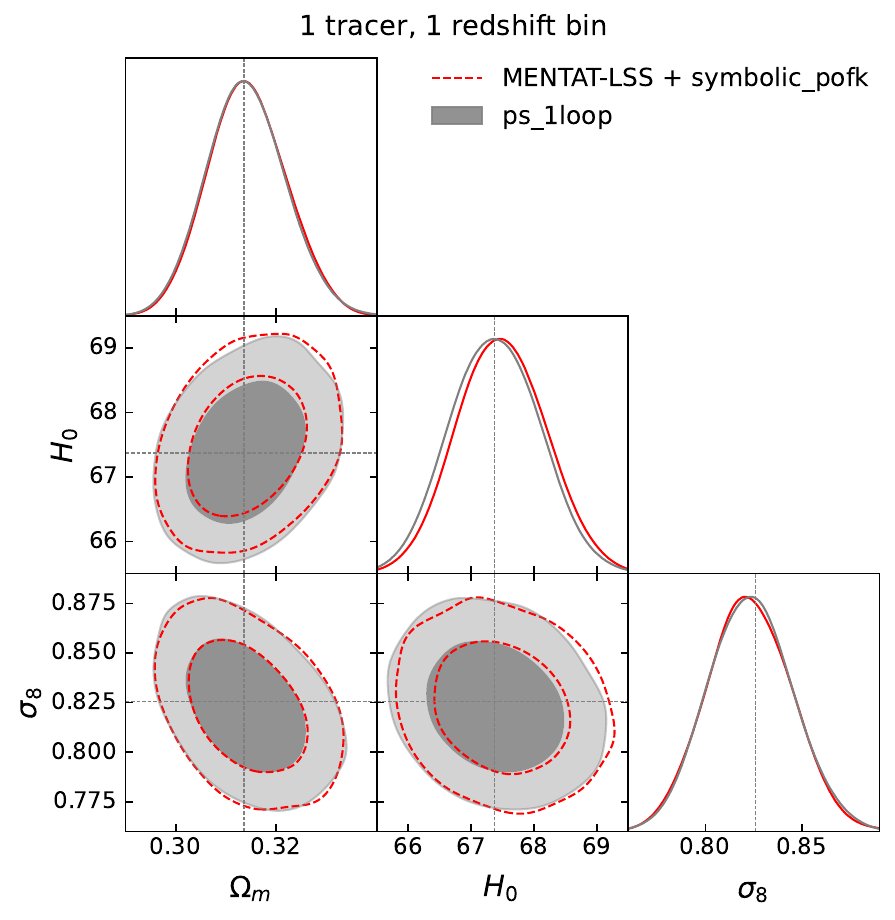}
    \caption{Parameter contours from simulated likelihood analyses with one tracer and one redshift bin, using pure EFT calculations (gray) and our power spectrum emulator (red). Dashed lines indicate the fiducial cosmology at which our simulated data vector is generated. Both contours lie almost directly on top of each-other, indicating our emulator recovers the expected posterior distribution.}
    \label{fig:1t_1z_contours}
\end{figure}

Our first analysis includes only a single tracer and redshift bin from our full setup. We chose the higher redshift bin $(0.4 < z < 0.6)$, and the lower redshift-uncertainty tracer bin. This reduced configuration provides a lower-dimensional validation of the emulator and likelihood pipeline before introducing the additional complexity of multi-tracer correlations. Because the data vector is simplified, we make this test more stringent by additionally varying the EFT counterterm and stochastic parameters $c_0, c_2, \tilde{c},$ and $P_\text{shot}$. We do so by including those power spectrum terms analytically with the following priors,

\begin{align}
    c_0, c_2 \sim \mathcal{N}(0, 30^2),\ \tilde{c} \sim \mathcal{N}(500, 500^2),\\ P_\textbf{shot} \sim \mathcal{N}(2000, 5000^2).
\end{align}

\noindent These ranges are taken from Ref. \cite{Wadekar-2020}, with the additional change that the stochastic parameter is instead centered on $1/n_g$ of the chosen tracer / redshift bin. We use \texttt{symbolic\_pofk}\footnote{\url{https://github.com/DeaglanBartlett/symbolic_pofk}} \cite{symbolic_pofk_linear} to quickly calculate the linear power spectrum required for calculating the counterterms. Overall, this setup has a total of twelve varied parameters.

\begin{table*}
    \centering
    \begin{tabular*}{ \linewidth}{@{\extracolsep{\fill}} ccccc}
        \hline
         Parameter & \multicolumn{2}{c}{One tracer / redshift bin} & \multicolumn{2}{c}{Two tracer / redshift bins}\\
         & best-fit shift & uncertainty difference & best-fit shift & uncertainty difference \\
         \hline \hline
         $\Delta \log{\mathcal{Z}}$ & \multicolumn{2}{c}{$-6.01 \times 10^{-2}$} & \multicolumn{2}{c}{$7.8 \times 10^{-1}$} \\
         \hline
         $h$ & $0.143\sigma$ & $0.73\%$ & $0.16\sigma$ & $2.28\%$ \\
         $\omega_b$ & $0.094\sigma$ & $1.63$\% & $0.21\sigma$ & $1.92\%$ \\
         $\omega_{cdm}$ & $0.096\sigma$ & $1.21$\% & $0.14\sigma$ & $1.04\%$\\
         $A_s$ & $0.071\sigma$ & $2.70$\% & $0.12\sigma$ & $1.00\%$ \\
         $n_s$ & $0.051\sigma$ & $0.64$\% & $0.10\sigma$ & $0.96\%$\\
         \hline
         $\Omega_m$ & $0.031\sigma$ & $0.92$\% & $0.019\sigma$ & $0.03\%$ \\
         $\sigma_8$ & $0.024\sigma$ & $0.02$\% & $0.072\sigma$ & $0.86\%$ \\
         \hline
         $\langle \theta_\text{cosmology}\rangle$ & $0.091\sigma$ & $1.38\%$ & $0.145 \sigma$ & $1.44\%$ \\
         $\langle \theta_\text{nuisance}\rangle$ & $0.039 \sigma$ & $2.23 \%$ & $0.054\sigma$ & $0.60 \%$\\
         \hline
    \end{tabular*}
    \caption{Absolute relative shifts of 1D best-fit values, and percent difference of 68\% confidence intervals, for varied parameters when comparing chains produced with our emulator vs \textit{ps\_1loop}. The values in column 2 and 3 correspond to Fig. \ref{fig:1t_1z_contours}, and columns 4 and 5 to Fig. \ref{fig:2t_2z_contours_smaller}. The last two rows show the average shifts split by cosmology and nuisance (galaxy bias, counterterm, and stochastic) parameters. We also display the change in Bayesian evidence between emulated and analytical analyses for each configuration.}
    \label{tab:shifts_table}
\end{table*}

Figure \ref{fig:1t_1z_contours} shows the marginalized posteriors from this first analysis, while Table \ref{tab:shifts_table} presents the 1D bestfit shifts for the cosmology parameters between our emulator and \textit{ps\_1loop} chains. By eye, we see good agreement between emulated and analytic contours. More quantitatively, the resulting 1D marginalized bestfit values have an average shift of $0.057 \sigma$, with $h$ having the largest deviation at $0.143 \sigma$. We also see the 1D error bars agree nicely with an average absolute percent difference of $2.2\%$, and $P_\text{shot}$ having the highest deviation at $5.1\%$. 

Next, we run analyses using all tracer and redshift bins in our setup. We first vary only directly emulated parameters, fixing all counterterms and stochastic terms to zero. Since each bin has independent nuisance parameters, we now have a total of seventeen varied parameters. 

\begin{figure}
    \centering
    \includegraphics[width=\linewidth]{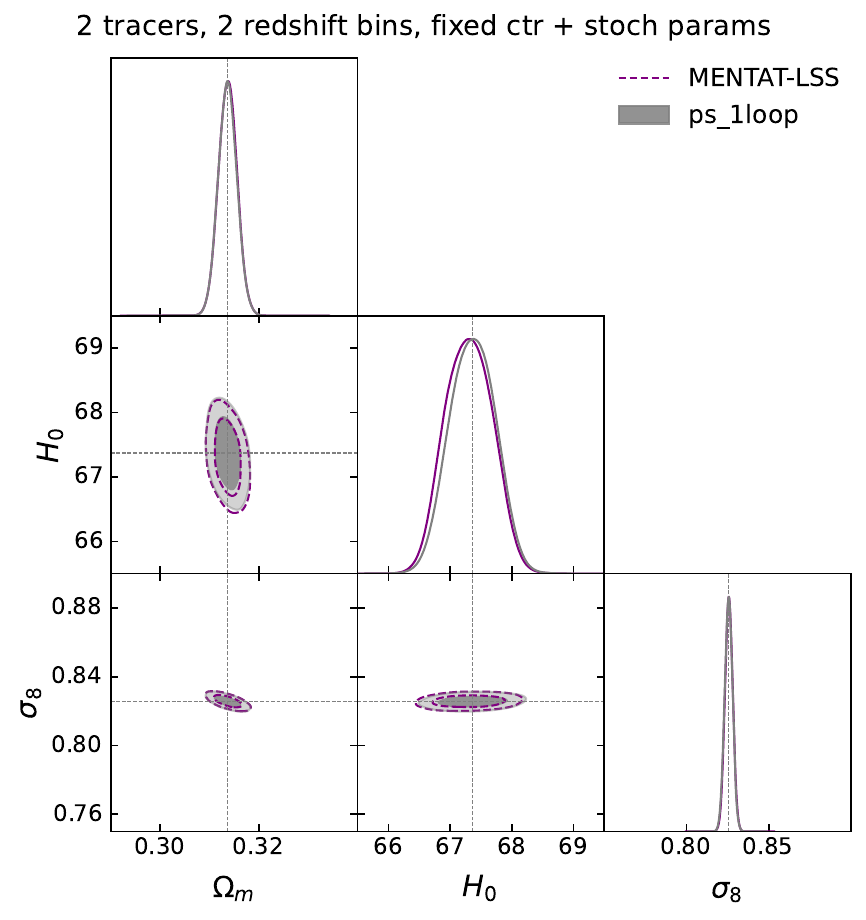}
    \caption{Parameter contours from simulated likelihood analyses with our full multi-tracer setup, using pure EFT calculations (gray) and our power spectrum emulator (purple). Axes have been scaled to match those in Fig. \ref{fig:1t_1z_contours}. Here, we set all (non-emulated) counterterm and stochastic parameters to 0. Similar to Fig. \ref{fig:1t_1z_contours}, we see good agreement between emulated and analytic contours.}
    \label{fig:2t_2z_contours_smaller}
\end{figure}

Figure \ref{fig:2t_2z_contours_smaller} shows the marginalized posteriors from this multi-tracer analysis, while Table \ref{tab:shifts_table} presents the 1D bestfit and absolute percent change in $68\%$ confidence intervals. Once again, we see good agreement between emulated and analytic contours, with an average 1D best-fit shift of $0.078\sigma$ and mean absolute uncertainty difference of $0.82 \%$. Looking closer, the nuisance parameters have slightly smaller deviations overall compared to the cosmology parameters, which indicates our emulator has an easier time handling those terms.

Finally, we perform a multi-tracer analysis in our full parameter space, where we now allow counterterm and stochastic parameters to vary using the same method as in Fig. \ref{fig:1t_1z_contours}. With these additional degrees of freedom, we now have have a total of thirty three varied parameters. 

\begin{figure}
    \centering
    \includegraphics[width=\linewidth]{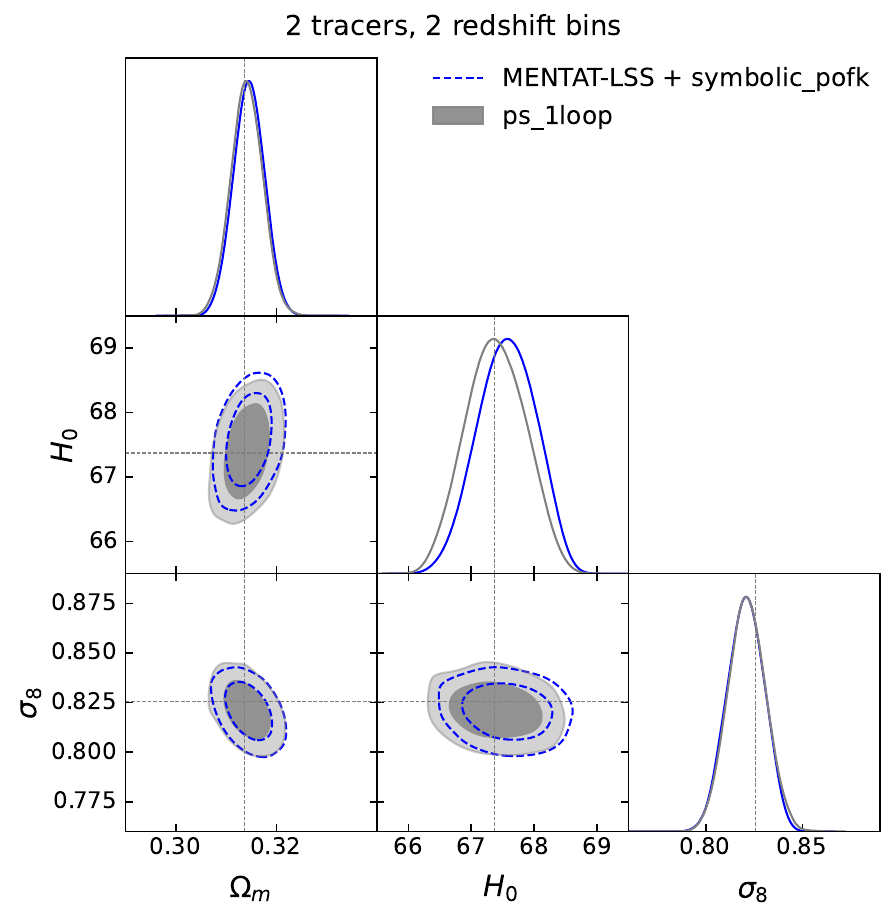}
    \caption{Same as Fig. \ref{fig:2t_2z_contours_smaller}, except now varying counterterm and stochastic parameters by additionally using \texttt{symbolic\_pofk}. Emulated and analytic contours have larger deviations compared to using MENTAT-LSS alone, indicating \texttt{symbolic\_pofk} may not be accurate enough to correctly vary these additional parameters.}
    \label{fig:2t_2z_contours}
\end{figure}

Figure \ref{fig:2t_2z_contours} shows the resulting contours of this more ambitious analysis. Overall, the average 1D best-fit shift is $0.078\sigma$. and the mean absolute uncertainty difference is $3.26 \%$. However, these statistics mask notably higher deviations in the varied cosmology parameters, which have an average best-fit shift of $0.29 \sigma$ and several parameters off by over $0.3 \sigma$. One plausible contributor to this discrepancy is the additional $0.5\%$ approximation error introduced by \texttt{symbolic\_pofk} \cite{symbolic_pofk_linear}, which can become more apparent with additional tracer / redshift bins that tighten constraints and break degeneracies. Indeed, while this error seems to be manageable for our single-tracer analysis, extending to larger multi-bin setups will likely require a more accurate treatment of the additional terms computed with \texttt{symbolic\_pofk} .

\subsection{Metric 2: Bayesian Evidence}

We also report the change in Bayesian evidence factor $\Delta \log{\mathcal{Z}}$ between emulated and analytical chains shown in Figs. \ref{fig:1t_1z_contours} and \ref{fig:2t_2z_contours_smaller}. This quantity is frequently used to compare the goodness of fit of different models and data-sets \cite{Bayesian-evidence-review, Bayesian-evidence-data}.

In our single-tracer analysis, we find $\Delta \log{\mathcal{Z}} =-6.01 \times 10^{-2}$ while for our multi-tracer analysis, $\Delta \log{\mathcal{Z}} = 7.8 \times 10^{-1}$. For both cases, the difference is much smaller than the true evidence calculated from our \textit{ps\_1loop} chains ($-33.03$ and $-86.28$ respectively), indicating that the emulator recovers the relative evidence (and thus model preference) to good accuracy.

\section{Summary \& Conclusion}
\label{sec:conclusion}

Predicting the galaxy power spectrum serves as a serious computational bottleneck when running simulated analyses for stage-IV surveys like SPHEREx. To enable faster impact studies in preparation for real data, this paper presents a fast method of generating galaxy power spectrum multipoles using a series of MLP and transformer-based neural networks. We use this neural network emulator to output the monopole and quadropole moments predicted by the EFTofLSS for two correlated tracer bins, and two independent redshift bins, all simultaneously. In this setup, our emulator can output power spectrum multipoles in $\sim 2.2$ms ($0.37$ms for one tracer / redshift bin combination) on a typical laptop, which is roughly $900$ times faster than the brute-force calculation ran on the same hardware. The code used to train and use this emulator is publicly available on GitHub and pip via the python package MENTAT-LSS (aka spherex-emu)\footnote{\url{https://github.com/jadamo/spherex_emu}}.

We optimize our emulator setup by individually varying several model and training hyperparameters, and using the median $\Delta \chi^2$ to determine how each one affected performance. We also test two different loss functions during this process, and find that both achieved the same overall accuracy. Next, we explore how performance changes when varying both the sampling strategy used to generate the training set, and the number of training set samples. Overall, our emulator performance follows a clear power law w.r.t. training set size, and uniform hypersphere sampling performs $\sim 2$ orders of magnitude better than using a Latin hypercube. These results indicate we can further improve performance if needed by increasing the amount of training data.

We tested the accuracy of our optimized emulator by computing the $\Delta \chi^2$ error between emulated and analytically-calculated power spectrum multipoles in the test set for each individual tracer / redshift bin, and for all bins together. All sub-networks perform well, with median $\Delta \chi_{(z, tt)}^2$ all at or below $10^{-2}$, and one bin achieving an impressive $10^{-3}$. The full emulator output achieves a median $\Delta \chi_{tot}^2 = 0.069$, with no observed parameter dependence. These results pass the target threshold $\Delta \chi^2 < 1$ used by both the DES collaboration, and other emulators of cosmological observables \cite{DES-Y3-Validation, LSST-emulator-1} .

Finally, we run simulated likelihood analyses with our emulator, using both a single tracer and redshift bin, and two of each. The resulting contours for the single-tracer analysis match with the expected distribution to great precision, with an average 1D best-fit shift of of $0.057\sigma$, a mean uncertainty percent difference of $2.23\%$, and Bayesian evidence change of $-6.01 \times 10^{-2}$. When keeping the stochastic and counterterm parameters fixed, our multi-tracer analysis results in similarly small deviations, with an average best-fit shift agreeing to $0.078\sigma$, a mean uncertainty percent difference of $0.82\%$, and Bayesian evidence change of $7.8 \times 10^{-1}$. Finally, we find that using \texttt{symbolic\_pofk} to calculate the counter-term contributions caused these deviations to increase noticeably in our multi-tracer setup, indicating we may need a more accurate method of varying those parameters for larger analyses.

This work corroborates recent efforts to emulate cosmological observables from both galaxy clustering and weak-lensing surveys (e.g. Refs. \cite{LSST-emulator-1, LSST-emulator-2, Derose-2022}). In particular, we obtain comparable performance to Ref. \cite{Derose-2022}, which presents a pca-based neural network to predict LPT-based galaxy power spectra in a similarly sized parameter space. They report 1D best-fit shifts $\lesssim 0.15\sigma$ and agreement in marginalized 1D uncertainties at the $\lesssim 10\%$ level, consistent with the level of performance we find here. Given the similar application, key differences in our works include the specific bias model used, the underlying network architecture, data pre-processing (we do not perform a PCA compression of the data vector), and our implementation that concurrently generates power spectra for multiple tracer and redshift bins.

To conclude, our emulator can demonstrably generate galaxy power spectra in a large enough input parameter space to run simulated analyses for SPHEREx. Overall, we consider this work as both a useful tool towards precise measurements with SPHEREx, and as a further demonstration of the promise of using machine learning to enhance simulated likelihood analyses.

\begin{acknowledgments}
We would like to thank Yosuke Kobayashi for both providing access and guidance to use his power spectrum modeling code. We acknowledge support from the SPHEREx project under a grant from the NASA/GODDARD Space Flight Center to the University of Arizona for SPHEREx L4 cosmology R\&D. The emulator developed in this paper used High Performance Computing (HPC) resources supported by the University of Arizona TRIF, UITS, and RDI and maintained by the UA Research Technologies department. 
\end{acknowledgments}

\bibliography{sources}

\end{document}